\documentclass[12pt]{article}
\usepackage{latexsym}
\usepackage{epsfig,amssymb,euscript}
\usepackage{amsmath}
\usepackage{array,calc,epsfig}
\usepackage{citesort}

\def\be{\begin{equation}}
\def\ee{\end{equation}}
\def\bea{\begin{eqnarray}}
\def\eea{\end{eqnarray}}

\numberwithin{equation}{section} %%
\usepackage[]{graphicx}

\def\calb         {{\cal B}}

\def\cald         {{\cal D}}
\def\cale         {{\cal E}}
\def\calf         {{\cal F}}

\def\calj         {{\cal J}}

\def\caln         {{\cal N}}

\def\calw         {{\cal W}}

\def\Re           {{\rm Re\hskip0.1em}}
\def\Im           {{\rm Im\hskip0.1em}}

\def\sqr#1#2{{\vcenter{\vbox{\hrule height.#2pt
 \hbox{\vrule width.#2pt height#1pt \kern#1pt \vrule width.#2pt}\hrule
 height.#2pt}}}}

\def\d{\text{d}}

\def\slashchar#1{\setbox0=\hbox{$#1$}           % set a box for #1
\dimen0=\wd0                                 % and get its size
\setbox1=\hbox{/} \dimen1=\wd1               % get siste of /
\ifdim\dimen0>\dimen1                        % #1 is bigger
\rlap{\hbox to \dimen0{\hfil/\hfil}}      % so center / in box
#1                                        % and print #1
\else                                        % / is bigger
\rlap{\hbox to \dimen1{\hfil$#1$\hfil}}   % so center #1
/                                         % and print /
\fi}

\begin{document}
\font\cmss=cmss10 \font\cmsss=cmss10 at 7pt
\leftline{\tt hep-th/0701093}

\vskip -0.5cm
\rightline{\small{\tt KUL-TF-07/02}}
%\rightline{\small{\tt Imperial/TP/050701}}

\vskip .7 cm
%\hfill IC/2004/ \vskip .1in \hfill CPHT \vskip .1in \hfill hep-th/0701093

\hfill
\vspace{18pt}
\begin{center}
{\Large \textbf{Supersymmetric D-branes on flux backgrounds}}
\end{center}

\vspace{6pt}
\begin{center}
{\large\textsl{Luca Martucci}}

\vspace{25pt}
\textit{ Institute for Theoretical Physics, K.U. Leuven,\\ Celestijnenlaan 200D, B-3001 Leuven, Belgium}\\  \vspace{6pt}
%\textit{\small $^b$ The Blackett Laboratory, Imperial College London, \\ Prince Consort Road, London SW7 2AZ, U.K.}
\end{center}

\vspace{12pt}

\begin{center}
\textbf{Abstract}
\end{center}

\vspace{4pt} {\small \noindent }
Several aspects concerning the physics of D-branes in Type II  flux compactifications preserving minimal $\caln=1$ supersymmetry in four dimensions are considered. It is shown how  these vacua are completely characterized in terms of properly defined generalized calibrations for D-branes and the relation with Generalized Complex Geometry is discussed.  General expressions for superpotentials and D-terms associated with the $\caln=1$ four-dimensional description of space-time filling D-branes are presented. The massless spectrum of calibrated D-branes can be characterized in terms of cohomology groups  of a differential complex canonically induced on the D-branes by the underlying generalized complex structure. 

\vspace{4cm}

\noindent {\em Contribution to the proceedings of the RTN project `Constituents, Fundamental Forces and Symmetries of the Universe' conference in Napoli, October 9 - 13, 2006.}

\vfill
\vskip 5.mm
\hrule width 5.cm
\vskip 2.mm
{\small
\noindent e-mail: luca.martucci@fys.kuleuven.be}

\newpage

%%%%%%%%%%%%%%%%%%%%%%%%%%%%%%%%%%%%%%%%
\section{Introduction}
\label{intro}

The introduction of fluxes on string/M-theory configurations with or without branes seems to be an unavoidable ingredient for obtaining many of the phenomenologically and/or theoretically interesting models. Here we will focus on Type II configurations preserving four-dimensional Poincar\'e invariance and minimal $\caln=1$ supersymmetry.  They constitute the natural generalization of the standard $\caln=2$ compactifications on Calabi-Yau manifolds and appear in the construction of realistic scenarios in string theory or in the context of the string/gauge theory correspondence for minimally supersymmetric and possibly confining gauge theories.  

D-branes can play a key role in most of the models. Thus the comprehension of some of their dynamical and geometrical properties, derivable from general considerations based on supersymmetry without the need of restricting on specific examples, can be obviously useful. The fluxless Calabi-Yau case constitutes a guiding example, where many general results can be obtained thanks to the geometrical structure  of  Calabi-Yau spaces, which are Ricci-flat K\"ahler manifolds, thus admitting a pair of compatible integrable complex and symplectic structures (for a review on D-branes on Calabi-Yau spaces see e.g. \cite{DCY}). However, by introducing fluxes on the internal six-dimensional space one generally loses the geometrical properties of Calabi-Yau spaces and all the results about D-branes on these spaces are no longer applicable in general. 

In this paper I will report on some of the results obtained in \cite{luca1,luca2,luca3}, in collaboration with P.~Smyth and P.~Koerber, where it was shown how the minimal supersymmetry preserved by the background still allows one to obtain substantial quantitative information about D-branes.\footnote{See also \cite{fernando1,fernando2}, where supersymmetric D-branes on Type II $SU(3)$-structure flux vacua are  studied and interesting effects originating from the presence of fluxes are discussed.}  The backgrounds we consider preserve four-dimensional Poincar\'e invariance. Thus, the ten dimensional space $\mathbb{R}^{1,3}\times M$ has  a  metric of the form
\bea
\d s^2=e^{2A(y)}\d x^\mu\d x_\mu +g_{mn}(y)\d y^m\d y^n\ ,
\eea                  
the Ramond-Ramond (RR) field-strengths split as $F=\text{vol}_4\wedge \tilde F + \hat F$, where $\tilde F$ and $\hat F$, as well as the Neveu-Schwarz (NS) $H$ field strength, have only internal legs. All the fields, including the dilaton $\Phi$, depend only on the internal coordinates $y^m$.  Minimal supersymmetry requires the existence of four independent background Killing spinors, expressed in terms of two ten-dimensional Majorana-Weyl spinors $\epsilon^{(1)}=\zeta_+\otimes \eta^{(1)}_+ + \ \text{c.c.}$ and  $\epsilon^{(2)}=\zeta_+\otimes \eta^{(2)}_\mp + \ \text{c.c.}$  in IIA/IIB. $\zeta$ is an arbitrary constant  Weyl spinor in four dimensions while $\eta^{(1)}$ and $\eta^{(2)}$ are Weyl spinors on the internal manifold that characterize the flux vacuum one is considering. In order to have static D-branes on a background of this kind, one has to slightly restrict this very general ansatz by imposing the necessary condition $||\eta^{(1)}||=||\eta^{(2)}||=|a(y)|^2$. Roughly speaking, this  condition identifies backgrounds that may be seen as the back-reaction of some supersymmetric D-brane configuration, and is also natural if one wants to introduce orientifolds. More precisely, as was discussed in \cite{luca1,luca2} and  we are going to review presently, this condition allows one to characterize this class of $\caln=1$ backgrounds as {\em D-calibrated} backgrounds.

\section{Generalized calibrations and supersymmetric D-branes}
\label{sec1}

We would like  to describe now the background supersymmetry conditions in a form that proves particularly useful when describing the geometrical and physical properties of D-branes. It turns out that the most suitable form is that found in \cite{gmpt}, in terms of (even/odd) complex polyforms $\Psi^\pm$ on the internal space, obtained by Clifford map from the bispinors $\eta^{(1)}_+\otimes \eta^{(2)}_\pm$. The polyforms $\Psi^\pm$ are pure spinors in the context of generalized complex geometry (GCG) \cite{hitchin,gualtieri} which, as we will see, is the most natural language to use in  $\caln=1$ flux models with (or without) D-branes \cite{gmpt,luca1,luca2,GCG,luca3}. $\Psi^\pm$ obey the compatibility conditions such that they reduce the $SO(6,6)$ structure group of $T_M\oplus T^\star_M$ to $SU(3)\times SU(3)$. In terms of the normalized pure spinors $\hat{\Psi}^\pm=(-8i/|a|^2)\Psi^\pm$, the supersymmetry conditions for D-calibrated backgrounds read 
\bea\label{backsusy}
\d_H(e^{4A-\Phi}\Re\hat\Psi_1)=e^{4A}\tilde F\quad,\quad \d_H(e^{2A-\Phi}\Im\hat\Psi_1)=0\quad,\quad \d_H (e^{3A-\Phi}\hat\Psi_2)=0\ ,
\eea 
where $\hat\Psi_1=\hat\Psi^{\mp}$ and $\hat\Psi_2=\hat\Psi^{\pm}$ in IIA/IIB, and  $\d_H\equiv \d +H\wedge$. Furthermore, one has to impose $\d (\log |a|^2)=\d A$, which can be derived directly from the $\caln=1$ algebra. The last equation in (\ref{backsusy}) tells us that the generalized complex structure $\calj_2$ defined by the pure spinor $\hat\Psi_2$ is integrable \cite{hitchin,gualtieri} or, more strictly, the internal space is a generalized Calabi-Yau \`a la Hitchin \cite{hitchin}.  On the other hand, the first equation in (\ref{backsusy}) implies that the RR-fluxes act as an obstruction to the integrability of the (almost) generalized complex structure $\calj_1$ defined by $\hat\Psi_1$. In the following, we will refer to $\hat\Psi_1$ and $\hat\Psi_2$ as the non-integrable and integrable pure spinor respectively. 

In order to introduce a supersymmetric D$p$-brane into our backgrounds, standard arguments from $\kappa$-symmetry imply that we must in principle check only a projection condition of the form
\bea\label{kappa}
&&\epsilon^{(2)}=\hat\Gamma_{\text{D}p} \epsilon^{(1)}\ .
\eea   
A key step is to express  (\ref{kappa}) in terms of appropriate generalized calibrations \cite{luca1}.  Ordinary calibrations in a metric space are closed $k$-forms which minimize the volume form induced on $k$-dimensional submanifolds \cite{HL} and are naturally present in supersymmetric flux-less vacua \cite{bbs}. The inclusion of background fluxes that couple minimally to the brane led to the introduction of generalized calibrations which minimize the brane's energy instead of volume \cite{gencalold}. However, in general flux configurations a definition of calibration which takes into account a possible non-trivial world-volume field-strength is needed \cite{koerber,luca1}. Following the definition of \cite{luca1}, given a ten-dimensional static background manifold of the form $\mathbb{R}\times X$ (where $\mathbb{R}$ denotes the time direction), a generalized calibration is given by a (real) polyform $\omega$ of definite parity on $X$ such that, for any generalized cycle $(\Sigma,\calf)$ \footnote{Using the definition of \cite{gualtieri}, a generalized complex submanifold $(\Sigma,\calf)$ is given by a submanifold $\Sigma$ and a field-strength $\calf$  on it such that $\d \calf=P_\Sigma[H]$, where $P_\Sigma[.]$ denotes the pull-back to $\Sigma$.} one has
\bea
\label{alg}&&P_\Sigma[\omega]\wedge e^\calf|_{\text{top}}\leq \cale(\Sigma,\calf)\quad\quad \qquad\quad\text{(algebraic condition)}\ ,\\
\label{diff}&&\d_H\omega=0\quad\quad\quad\quad\qquad \qquad\qquad\qquad\text{(differential condition)}\ .
\eea      
In the first condition (\ref{alg})  $\cale(\Sigma,\calf)$ denotes the standard DBI+CS energy density  of the D-brane wrapping the generalized cycle $(\Sigma,\calf)$ and $|_{\text{top}}$ selects the top form on $\Sigma$.\footnote{See \cite{witt} for a study of the transformation properties of this kind of calibrations under T-duality.}

For the $\caln=1$ backgrounds we are interested in $X=\mathbb{R}^3\times M$ and we expect three kinds of possible supersymmetric D-branes: filling all three (space-time filling),  one (strings) or two (domain walls) flat directions in $\mathbb{R}^3$  and wrapping some internal generalized cycle $(\Sigma,\calf)$. Indeed, as the name  suggests, $\caln=1$ D-calibrated backgrounds described above admit, and are in fact characterized by, generalized calibrations on $M$ for each of these configurations \cite{luca1}
\bea\label{dcal}
\omega^{\text{(4d)}}=e^{4A}(e^{-\Phi}\Re\hat\Psi_1-\tilde C)\ ,\ \omega^{\text{(string)}}=e^{2A-\Phi}\Im\hat\Psi_1\ ,\ \omega^{\text{(DW)}}=e^{3A-\Phi}\Re(e^{i\theta}\hat\Psi_1)\ ,
\eea  
where $\tilde C$ is such that $\tilde F=e^{-4A}\d_H(e^{4A}\tilde C)$ and $\theta$ is the (in principle arbitrary) phase specifying the one-half of supersymmetry preserved by the domain wall. Note that imposing (\ref{diff}) for each of the calibrations (\ref{dcal}) (and for any $\theta$) one exactly obtains the background  supersymmetry conditions (\ref{backsusy}). In particular the non-integrability of the pure spinor $\hat\Psi_1$ is in fact required by the integrability of the calibration $\omega^{\text{(4d)}}$.

The supersymmetry condition (\ref{kappa}) for a D-brane  can be written as 
\bea\label{calcond}
P_\Sigma[\omega]\wedge e^\calf|_{\text{top}}= \cale(\Sigma,\calf)
\eea
i.e.\ the D-brane must wrap a calibrated generalized cycle $(\Sigma,\calf)$. By generalizing the standard arguments of calibrated geometry \cite{HL,gencalold},  if $(\Sigma^\prime,\calf^\prime)$ is continuously connected to a supersymmetric/calibrated generalized cycle $(\Sigma,\calf)$, then $E(\Sigma,\calf)\leq E(\Sigma^\prime,\calf^\prime)$, that is calibrated generalized cycles are energy minimizing inside their (generalized) homology class \cite{luca1}.

\section{Superpotentials and D-terms}
\label{sec2}

Let us now focus on space-time filling D-branes. The calibration condition (\ref{calcond}) is equivalent, up to an appropriate choice of orientation, to the pair of conditions\footnote{Calibrated  strings are alternatively defined by the conditions $P_\Sigma[\Re \hat\Psi_1]\wedge e^\calf|_{\text{top}}=0$ and $P_\Sigma[(\iota_m+g_{mn}\d y^n\wedge)\hat\Psi_1]\wedge e^\calf|_{\text{top}}=0$, while domain walls by $P_\Sigma[\Im (e^{i\theta}\hat\Psi_2)]\wedge e^\calf|_{\text{top}}=0$ and $P_\Sigma[(\iota_m+g_{mn}\d y^n\wedge)\hat\Psi_1]\wedge e^\calf|_{\text{top}}=0$.}
\bea\label{altcal}
\cald(\Sigma,\calf)&=&P_\Sigma[e^{2A-\Phi}\Im \hat\Psi_1]\wedge e^\calf|_{\text{top}}=0\quad\quad\qquad\qquad\quad\quad \text{(D-flatness)}\ ,\cr
\calw_m(\Sigma,\calf)&=&P_\Sigma[e^{3A-\Phi}(\iota_m+g_{mn}\d y^n\wedge)\hat\Psi_2]\wedge e^\calf|_{\text{top}}=0\quad\ \quad \text{(F-flatness)}\ .
\eea 
By directly expanding the DBI+CS D-brane action around a supersymmetric configuration, it is possible to show \cite{luca2} that $\cald(\Sigma,\calf)$ and $\calw_m(\Sigma,\calf)$ correspond to the D-term  and the F-term respectively, in the four-dimensional description of the space-time filling D-brane wrapping the internal cycle $(\Sigma,\calf)$. Moreover, it is possible to introduce a superpotential $\calw$ that, differentiated, generates the F-term $\calw_m$.  Indeed, choose a fixed generalized cycle $(\Sigma_0,\calf_0)$, so that there exists a generalized chain $(\calb,\tilde\calf)$ interpolating between $(\Sigma_0,\calf_0)$ and $(\Sigma,\calf)$, i.e. $\partial\calb=\Sigma-\Sigma_0$, $\tilde\calf|_{\Sigma}=\calf$ and  $\tilde\calf|_{\Sigma_0}=\calf_0$. Then, the superpotential is given by \cite{luca2}
\bea\label{supot}
\calw(\Sigma,\calf)=\int_\calb P_\calb[e^{3A-\Phi}\hat\Psi_2]\wedge e^{\tilde\calf}\ .
\eea
Note that the superpotential (\ref{supot}) depends only on the  integrable pure spinor $\hat\Psi_2$. Indeed, it can be defined on any generalized Calabi-Yau, ignoring the existence of the additional structure given by $\hat\Psi_1$. It can be shown \cite{luca2,luca3} that $\calw(\Sigma,\calf)$ is extremized at {\em generalized complex cycles} as defined in \cite{gualtieri}, which provide the natural generalization in generalized complex geometry of  holomorphic or Lagrangian cycles in  ordinary complex and symplectic geometry.\footnote{It is possible to show  that the superpotential (\ref{supot}) is holomorphic with respect to a properly defined (almost) complex structure on the configuration space of generalized cycles \cite{luca2}.} The superpotential  (\ref{supot}) may also  be derived from a physical argument. Suppose that $(\Sigma_0,\calf_0)$ and $(\Sigma,\calf)$ are two supersymmetric cycles and  consider a domain wall interpolating between the two configurations which can be obtained by wrapping a D-brane on $(\calb,\tilde\calf)$. From standard arguments in four-dimensional supersymmetric field theory, the supersymmetric domain wall tension is given by $T_{\text{DW}}=|\Delta\calw|$. On the other hand, from (\ref{calcond}) with $\omega^{\text{(DW)}}$ in (\ref{dcal}), the tension of the supersymmetric domain wall D-brane is exactly given by the right-hand side of (\ref{supot}).  

The D-term $\cald(\Sigma,\calf)$ depends only on the imaginary part of the non-integrable pure spinor $\hat\Psi_1$ \footnote{A formal symplectic structure  can be introduced into the configuration space so that the D-flatness condition $\cald(\Sigma,\calf)$ can be interpreted as the vanishing of the moment map generating the world-volume gauge transformations \cite{luca2}.} and, as the explicit form of $\omega^{\text{(string)}}$ in (\ref{dcal}) suggests, there should be a relation with D-brane strings in four dimensions. Indeed, one can observe that $\xi=\int_\Sigma\cald(\Sigma,\calf)$ is constant under continuous deformations of $(\Sigma,\calf)$  and can be identified as  the constant Fayet-Iliopoulos (FI) term for the residual $U(1)$ gauge symmetry in the KK reduction to four dimensions. Considering a D$\bar{\text{D}}p$-brane system wrapping $(\Sigma,\calf)$ such that $\xi\neq 0$, there is a complex tachyon field which is charged under the relative $U(1)$ which has FI-tem $\xi$. Thus, in the four-dimensional field theory description, one can have a standard BPS vortex solution with tension $T_{\text{string}}=2\pi\xi$. In the D-brane picture, the string should be given by the D$(p-2)$-brane surviving the tachyon condensation in the vortex solution and wrapping $(\Sigma,\calf)$. Thus, using the calibration condition (\ref{calcond}) with $\omega^{\text{(string)}}$ and taking into account the appropriate dimensionful factors, the tension of the BPS string is exactly given by $2\pi\int_\Sigma\omega^{\text{(string)}}\wedge e^\calf= 2\pi\xi$, thus reproducing the four-dimensional field theory result.         

For a detailed discussion of the explicit form of  superpotentials and D-terms in the subcase of $SU(3)$-structure backgrounds, showing how the above formulas for $\cald$ and $\calw$ include and generalize previous results present in the literature, see \cite{luca2} and references therein.

\section{Deformations of supersymmetric space-time filling D-branes}
\label{sec3}

The infinitesimal deformations of a generalized cycle $(\Sigma,\calf)$ are described by the sections of the generalized normal bundle  $\caln_{(\Sigma,\calf)}\equiv (T_M\oplus T^\star_M)|_\Sigma/T_{(\Sigma,\calf)}$ \cite{luca2}, where $T_{(\Sigma,\calf)}$ is the generalized tangent bundle as defined in \cite{gualtieri}.  By definition, $(\Sigma,\calf)$ is a generalized complex cycle with respect to an integrable generalized complex structure $\calj:T_M\oplus T^\star_M\rightarrow T_M\oplus T^\star_M$ if $T_{(\Sigma,\calf)}$ is  stable under $\calj$  \cite{gualtieri}. Thus, on generalized complex cycles, $\calj$ naturally induces a complex structure on $\caln_{(\Sigma,\calf)}$. One can thus split $\caln_{(\Sigma,\calf)}\otimes \mathbb{C}=\caln_{(\Sigma,\calf)}^{1,0}\oplus \caln_{(\Sigma,\calf)}^{0,1}$ and identify a (real) section $\mathbb{X}$ of $\caln_{(\Sigma,\calf)}$ with its $(0,1)$-component $\mathbb{X}^{0,1}\in\Gamma(\caln_{(\Sigma,\calf)}^{0,1})$. 

Supersymmetric space-time filling D-branes wrap generalized cycles $(\Sigma,\calf)$ calibrated  by $\omega^{\text{(4d)}}$. As we have explained in section  \ref{sec2}, this corresponds to requiring that $(\Sigma,\calf)$ must obey the D and F-flatness conditions (\ref{altcal}). In particular, the F-flatness is equivalent to the condition that $(\Sigma,\calf)$  is generalized complex with respect to $\calj_2$. Thus we can use $\calj_2$ to define  $\caln_{(\Sigma,\calf)}^{0,1}$ and use its sections to describe the deformations of the calibrated generalized cycle. $\calj_2$ also defines a natural  complexification of the world-volume $U(1)$ gauge symmetry  and the F-flatness condition is invariant under the full complexified gauge symmetry. The D-flatness condition provides a gauge fixing for the imaginary extension of the real gauge symmetry. Thus calibrated generalized cycles are identified (up to stability issues) by equivalence classes of generalized complex cycles related by a complexified gauge transformation. 

In \cite{luca3} it was shown that infinitesimal deformations preserving the calibration condition are described by a cohomology group on $(\Sigma,\calf)$ depending only on the background integrable generalized complex structure $\calj_2$. To explain this in some more detail, let us recall that it is possible to introduce a   differential 
\bea
\d_{(\Sigma,\calf)}:\Gamma(\Lambda^k {\cal N}^{0,1}_{(\Sigma,\calf)})\rightarrow\Gamma(\Lambda^{k+1} {\cal N}^{0,1}_{(\Sigma,\calf)})\ .
\eea
which makes $\Lambda^\bullet {\cal N}^{0,1}_{(\Sigma,\calf)}$
 a differential complex.\footnote{Using the natural indefinite metric on $T_M\oplus T^\star_M$, the differential complex $(\Lambda^\bullet {\cal N}^{0,1}_{(\Sigma,\calf)},\d_{(\Sigma,\calf)})$ can be shown \cite{luca3} to be isomorphic to the Lie algebroid differential complex introduced in \cite{kapu}, whose cohomology groups describe the BRST spectrum of topological branes on generalized complex spaces \cite{kapu2}. Furthermore $(\Lambda^\bullet {\cal N}^{0,1}_{(\Sigma,\calf)},\d_{(\Sigma,\calf)})$ is elliptic \cite{luca3}.} Now, if $\mathbb{X}^{0,1}$ is a section of ${\cal N}^{0,1}_{(\Sigma,\calf)}$, then it preserves the F-flatness condition in (\ref{altcal}) if  $\d_{(\Sigma,\calf)}\mathbb{X}^{0,1}=0$. 
Furthermore, the complexified gauge symmetry acts as 
$\mathbb{X}^{0,1}\rightarrow \mathbb{X}^{0,1}+\d_{(\Sigma,\calf)}\lambda$, where $\lambda$ is any complex function on $\Sigma$.

On the other hand, from the DBI action, one can introduce a metric  depending on $\hat\Psi_1$ on the sections of $\Lambda^k {\cal N}^{0,1}_{(\Sigma,\calf)}$ and thus a codifferential
\bea
\d^\dagger_{(\Sigma,\calf)}:\Gamma(\Lambda^k {\cal N}^{0,1}_{(\Sigma,\calf)})\rightarrow\Gamma(\Lambda^{k-1} {\cal N}^{0,1}_{(\Sigma,\calf)})\ .
\eea
Then it is possible to see that  $\mathbb{X}^{0,1}$ preserves the D-flatness condition if $\Im(\d^\dagger_{(\Sigma,\calf)}\mathbb{X}^{0,1})=0$. Observing that $\Re(\d^\dagger_{(\Sigma,\calf)}\mathbb{X}^{0,1})=0$ provides a fixing for the real gauge symmetry, one can impose $\d^\dagger_{(\Sigma,\calf)}\mathbb{X}^{0,1}=0$ for physically distinguishable deformations, arriving at the result that infinitesimal deformations preserving the calibration condition are given by sections of ${\cal N}^{0,1}_{(\Sigma,\calf)}$ that are harmonic with respect to the Laplacian operator $\Delta_{(\Sigma,\calf)}=\d_{(\Sigma,\calf)}\d^\dagger_{(\Sigma,\calf)}+\d^\dagger_{(\Sigma,\calf)}\d_{(\Sigma,\calf)}$. Alternatively, as already anticipated, they can be identified with the element of the first cohomology group  
\bea
H^1(\Sigma,\calf)\equiv \text{ker}\, (\d_{(\Sigma,\calf)}|_1)/\text{im}\, (\d_{(\Sigma,\calf)}|_{0})\ .
\eea
Several examples have been presented in \cite{luca3}, where known results on holomorphic and special Lagrangian cycles in Calabi-Yau spaces are reproduced in the above unifying  formulation and the effect of world-volume and background fluxes is discussed in detail. Furthermore, some non-trivial features arising from a   `genuine' generalized complex space are analyzed by considering the simple but non-trivial case of a D3-brane on a point where the `type' \cite{gualtieri} of the generalized complex IIB background  jumps from one to three.  

Finally, let us stress that $H^1(\Sigma,\calf)$ represents the allowed first-order infinitesimal deformations, or the massless chiral fields $\phi^i$ in the four-dimensional description. The integrability properties for extending these infinitesimal deformations to finite ones remain to be studied. From a more physical point of view, this problem should be encoded in the effective superpotential $\calw_{\text{eff}}(\phi)$ for the massless fields, that may be obtainable from the geometrical superpotential (\ref{supot}) by integrating out the massive modes.

\vspace{1cm}

\section*{Acknowledgements}

I would like to thank P.~Koerber and P.~Smyth for their collaboration on the results presented in this paper. Thanks also to F.~Denef, M.~Petrini, A.~Tomasiello and A. Van Proeyen for useful discussions. This work is partially  supported by  the Federal Office for Scientific, Technical and Cultural Affairs through the ``Interuniversity Attraction Poles Programme -- Belgian Science Policy" P5/27 and by the European Community's Human Potential Programme under contract
MRTN-CT-2004-005104 `Constituents, fundamental forces and symmetries of the universe'.

%%%%%%%%%%%%%%%%%%%%%%%%%%%%%%%%%%%%%%%%%%%%%%%%%

\end{document}